\documentclass[largeformat]{interact}

\usepackage{epstopdf}
\usepackage[caption=false]{subfig}
\usepackage{enumitem}
\usepackage[dvipsnames]{xcolor}
\usepackage[numbers,sort&compress,merge]{natbib}
\bibpunct[, ]{[}{]}{,}{n}{,}{,}
\renewcommand\bibfont{\fontsize{10}{12}\selectfont}

\theoremstyle{plain}

\theoremstyle{definition}

\theoremstyle{remark}

\begin{document}

\articletype{}

\title{Macroscopic production of spin-polarized hydrogen atoms from the IR-excitation and photodissociation of molecular beams}

\author{
\name{C.~S. Kannis\textsuperscript{a,b}, J. Suarez\textsuperscript{c} and T.~P. Rakitzis\textsuperscript{b,d}\thanks{CONTACT T.~P. Rakitzis. Email: ptr@iesl.forth.gr}}
\affil{\textsuperscript{a}Institute for Nuclear Physics, Forschungszentrum J\"{u}lich, 52425 J\"ulich, Germany; \textsuperscript{b}University of Crete, Department of Physics, Herakleio, Greece;\textsuperscript{c}Departamento de Quimica, Modulo 13, Universidad Autonoma de Madrid, Cantoblanco 28049, Madrid, Spain;\textsuperscript{d}Foundation for Research and Technology Hellas, Institute of Electronic Structure and Laser, N. Plastira 100, Heraklion, Crete, Greece, GR-71110}
}

\maketitle

\begin{abstract}
We describe methods for the production of spin-polarized H and D atoms from the IR-excitation and photodissociation of molecular beams of HBr, HI, and NH\textsubscript{3} isotopes, including optical excitation schemes with partial hyperfine resolution. We discuss the extent to which the production rates may approach the IR-laser production rates of ${\rm 10^{21}\, photons\, s^{-1}}$, and how the production rates of conventional methods of ${\sim}{\rm 10^{17} \, s^{-1}}$ may be surpassed significantly.
\end{abstract}


\section{Introduction}

Nuclear spin-polarized hydrogen (SPH) and deuterium (SPD) atoms are used as polarized targets in atomic, molecular, nuclear, and particle physics~\cite{steffens,redsun}. Also, SPH and SPD atoms can be recombined at surfaces, to produce highly nuclear-spin-polarized H\textsubscript{2} isotopes~\cite{engels,engels2}. Pure SPH is conventionally produced by atomic-beam spin-separation methods~\cite{steffens}, and via spin-exchange optical pumping~\cite{poelker,clasie,redsun}. In both cases, the production rates of only up to $10^{17}$ SPH${\rm\, s^{-1}}$ have been demonstrated~\cite{nass}. Higher production rates are desirable, to increase the signals in collision experiments with small cross sections, but also to increase the production rate of polarized H\textsubscript{2} isotopes. It is known that the cross sections of the fusion reactions D$+$T and D$+$\textsuperscript{3}He are increased by 50\% when the nuclei are fully polarized~\cite{hupin}, and there is the potential to increase the efficiency of a fusion reactor by 75\%~\cite{temporal}, using polarized nuclei; however a fusion reactor will need about $10^{21}$ SPD${\rm\, s^{-1}}$~\cite{grigoryev,kulsrud,moir} which is about 4 orders of magnitude higher than the capabilities of the production rates of conventional methods.

The production of spin-polarized atoms, from the IR excitation and photodissociation of molecular beams, was proposed and demonstrated using pulsed lasers~\cite{rakitzis,rubio,sofikitis,sofikitis2}. The IR-excitation with circularly polarized light produces rotationally polarized molecules. The rotational polarization is then transferred to the nuclear spin via the hyperfine interaction; when the nuclear spin polarization is maximized, the molecules are photodissociated, to produce spin-polarized atoms. However, the spin-polarized atom production rate is limited by the production rate of IR photons. Conventional tunable pulsed IR lasers produce about ${\rm 10^{17} \, photons \, s^{-1}}$, so that atomic-beam separation methods cannot be easily surpassed using this method. In addition, no demonstrations of the production of SPH were achieved using this method, either because the rotational polarization was transferred to nuclei with stronger hyperfine coupling (such as to Cl nuclei, in the case of rovibrational polarization of HCl molecules~\cite{sofikitis,sofikitis2}), or because the polarization transfer time was very long (about 300 ${\rm \mu s}$ to reach 100\% polarization, in the case of acetylene) and molecular depolarization can occur in the molecular beam.

Recently, tabletop high-power cw IR lasers have become available, that are tunable with narrow linewidths, and can produce  ${\rm10^{21}\, photons \, s^{-1}}$~\cite{mirov}. This development allows the possibility of SPH production rates that can approach ${\rm 10^{21} \, s^{-1}}$. Very recently, we have given proposals for the production of spin-polarized H\textsubscript{2} molecules from the excitation and photodissociation of the special case of formaldehyde~\cite{mnhfp}. Here, we give detailed proposals on how large atomic SPH production rates can be achieved using the IR-excitation of molecular beams, while also giving solutions to the drawbacks mentioned above. We choose HBr, HI, and NH\textsubscript{3} isotopes, because these molecules have large IR and photodissociation cross sections (though these methods can be generalized to other molecules as well). Photodissociation cross secions on the non-rotating diatomic molecules HCl and HI were obtained from different initial vibrational states by
means of wave packet propagation. Specifically, in section~\ref{steps} we give an outline of the steps needed to produce and use large fluxes of SPH; in section~\ref{hcldcl} we describe the IR-excitation of HBr, DBr, HI, and DI with partial hyperfine resolution, allowing highly spin-polarized H and D; in section~\ref{dcl100pc} we describe a method for producing 100\% polarized D atoms; finally, in section~\ref{ammonia} we describe the production of SPH and SPD from the IR-excitation of ammonia isotopes.

\section{Description of the method}\label{steps}

In this section, we give details on the production of highly polarized H or D from the IR-excitation and photodissociation of HBr, HI, and NH\textsubscript{3} isotopes. The steps are described generally, but specific numbers are given for the case of HBr (for production rates ${\sim}10^{20}$ ${\rm SPH \,  s^{-1}}$):

\begin{figure}
	\centering
	\includegraphics*[width=.6\textwidth]{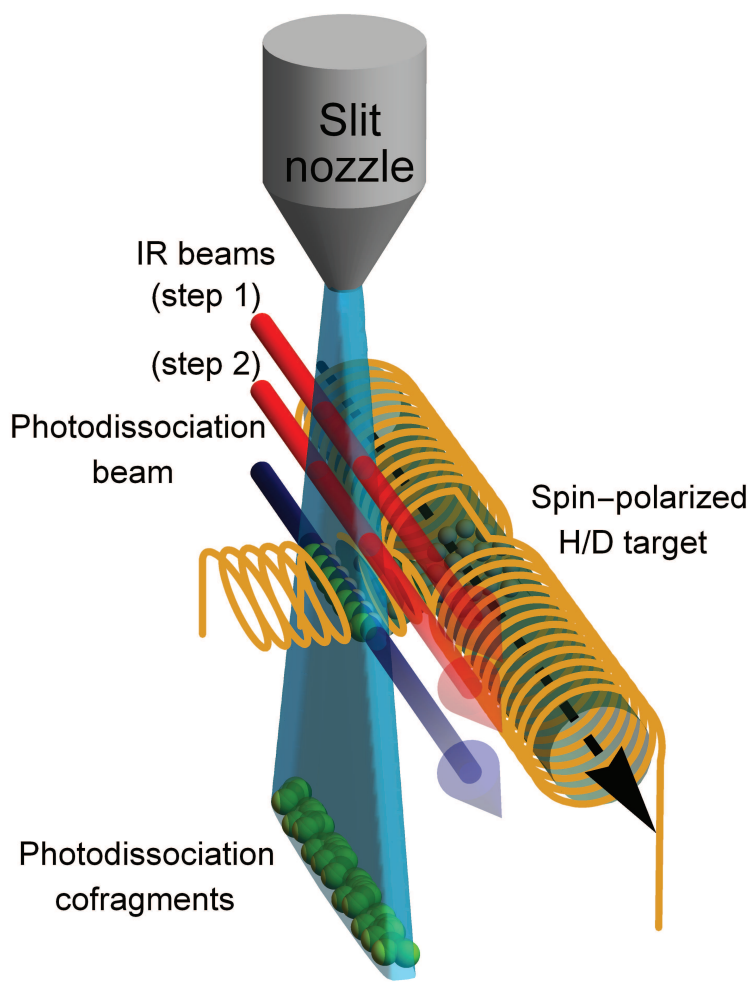}
	\caption{Experimental setup: supersonic expansion of HBr gas, followed by IR excitation, time evolution for time $t=t_{max}$, and then photodissociation.} \label{fig:setup}
\end{figure}

\begin{enumerate}[noitemsep,topsep=0pt,parsep=0pt,partopsep=0pt,leftmargin=0pt,itemindent=3em]
	\item \textit{Slit nozzle supersonic expansion.} Produces a ${\rm 10\,cm\times 1\, cm}$ jet (Fig.~\ref{fig:setup}), with HBr density of $2{\times} 10^{15}$ cm\textsuperscript{-3} (seeded in He) with beam velocity of ${\rm 2{\times} 10^5 \, cm \, s^{-1}}$, translational and rotational temperature of a few K, thus cooling ${>}90\%$ of the HBr to the ground $J=0$ state, and producing narrow infrared linewidths (${\sim 50}$ MHz have been demonstrated with slit nozzle expansions~\cite{schulder}).
	\item \textit{IR excitation (step 1).} Transition from the ground state $|\nu=0 ,\, J=0,\, m_J =0 \rangle$ to $|\nu^\prime ,\, J^\prime,\, m_{J^\prime} =J^\prime \rangle$, with n $\sigma_{+}$ photons from cw IR lasers (details given in later sections). The absorption cross sections are greater than $3{\times}10^{-16}$ cm\textsuperscript{2}, and the column density of the ground state is ${\sim}4{\times} 10^{16}$ cm\textsuperscript{-2} (including reflecting back the IR beams) so that at least ${\rm 10^{20} \, s^{-1}}$ molecules are transferred to the final state, from using ${\rm 10^{21} \, s^{-1}}$ IR photons for each transition step (we note that the 1 MHz laser linewidth is much narrower than the transition linewidth). These transitions can either be performed by nearly saturating each step (absorbing nearly 50\% of the photons in each step), or by using STIRAP~\cite{bergmann,vitanov,bergmann2,mukherjee}, for which population-transfer efficiency is typically ${>}90\%$, however about 10\% of the IR photons are absorbed (also note that for STIRAP the molecules must encounter the IR beams in reverse order). We note that that the effect of AC-Stark shifting of the hyperfine states, and the possibility of interfering with the m-state selectivity of the excitation scheme, still needs to be investigated.
	\item \textit{Hyperfine polarization beating.} The rotational polarization is transferred to nuclear polarization (details given in Sections~\ref{hcldcl}-\ref{ammonia}), until the nuclear polarization is maximized at $t=t_{max}$ (hyperfine beatings under similar conditions have been demonstrated for HCl, HD, D\textsubscript{2}~\cite{sofikitis,bartlett,bartlettd2}). The overlapping IR beams of each step are focused down to ${\rm {\sim} 0.5 \,mm}$ along the overlap direction (with cylindrical lenses), so that the positional uncertainty of the formation final states is ${\rm {\sim} 0.5 \, mm}$. The positional blurring corresponds to a blurring of ${\sim} 5 \%$ in the polarization beating time of ${\rm {\sim} 5 \, \mu s}$ (assuming a beam velocity of ${\rm 2 \times 10^5 \, cm\, s^{-1} }$), which will reduce the polarization negligibly (${<}1\%$).  
	\item \textit{Hyperfine beating stopped.} At $t=t_{max}$, the beam enters a magnetic field of 1 mT, which stops the hyperfine beating~\cite{kannis}. The field gradually increases to about 1 T, to prevent exchange of polarization with the SPH electron after photodissociation~\cite{engels}. 
	\item \textit{IR excitation (step 2) and photodissociation.} For cases where necessary, a further IR transition selectively excites only states that are 100\% nuclear-spin polarized. Finally, a cw laser selectively photodissociates only the highest excited states that are maximally spin polarized (as the photodissociation curves of the highest states are sufficiently shifted to the red).  For HBr column density of ${\sim}10^{16}$ cm\textsuperscript{-2} and a photodissociation cross section of ${\sim}10^{-19}$ cm\textsuperscript{2} at 266 nm (Fig.~\ref{fig:cross}), at least 10\% of the molecules are photodissociated, for laser fluxes of ${\rm {\sim}10^{20} \, photons \, cm^{-2} \,  s^{-1}}$. A buildup cavity with finesse of up to 1000~\cite{cooper} must be implemented for the 100 W photodissociation laser at 266 nm~\cite{kojima}. Use of HI/DI can shift the photodissociation to 355 nm where there are more powerful lasers. In addition, the larger photodissociation cross sections for HI and NH\textsubscript{3} ($10^{-18}$ cm\textsuperscript{2} and $10^{-17}$ cm\textsuperscript{2}, respectively) require more modest buildup cavities (with lower finesse).
	\item \textit{Target tube.} The SPH/SPD will have a recoil velocity much faster than the molecular beam speed, and will recoil to the target tube (Fig.~\ref{fig:setup}). In contrast, the much heavier photodissociation cofragments (Br, I, NH\textsubscript{2}) will have much slower recoils speeds, and will follow the molecular beam away from the target tube (and can be trapped separately).
\end{enumerate}

The product of the efficiencies of the above steps 1-6 is expected to exceed ${\sim}1\%$. Therefore, $10^{21}$ IR photons s\textsuperscript{-1} may produce at least $10^{19}$ SPH s\textsuperscript{-1} (though we note that the efficiencies of the IR-excitation steps and UV-photodissociation steps are yet to be demonstrated). We note that the atomic beam spin-separation method has been limited to beam areas of about 1 cm\textsuperscript{2}, and separation times of about 1 ms. For the IR-excitation method, the lack of need to spatially separate spins, and the faster polarization times of order 10 ${\rm \mu s}$, allow the molecular beam to be much bigger, and the molecular densities to be much higher. Together, these factors may allow the IR-excitation method to surpass the production rates of the conventional methods.

\section{Polarization via partial hyperfine resolution}\label{hcldcl}
\subsection{HBr}
\begin{figure}
	\centering
	\subfloat[IR-excitation steps of H\textsuperscript{79}Br presented in the partial hyperfine resolution. At $t=0$, the $|\nu^{\prime\prime} ,\, J=2,\, F_i = 1/2 \rangle$ state is populated and the molecule is prepared in the $m_{F_i} = + 1/2$ substate. ]{%
		\resizebox*{9.5cm}{!}{\includegraphics{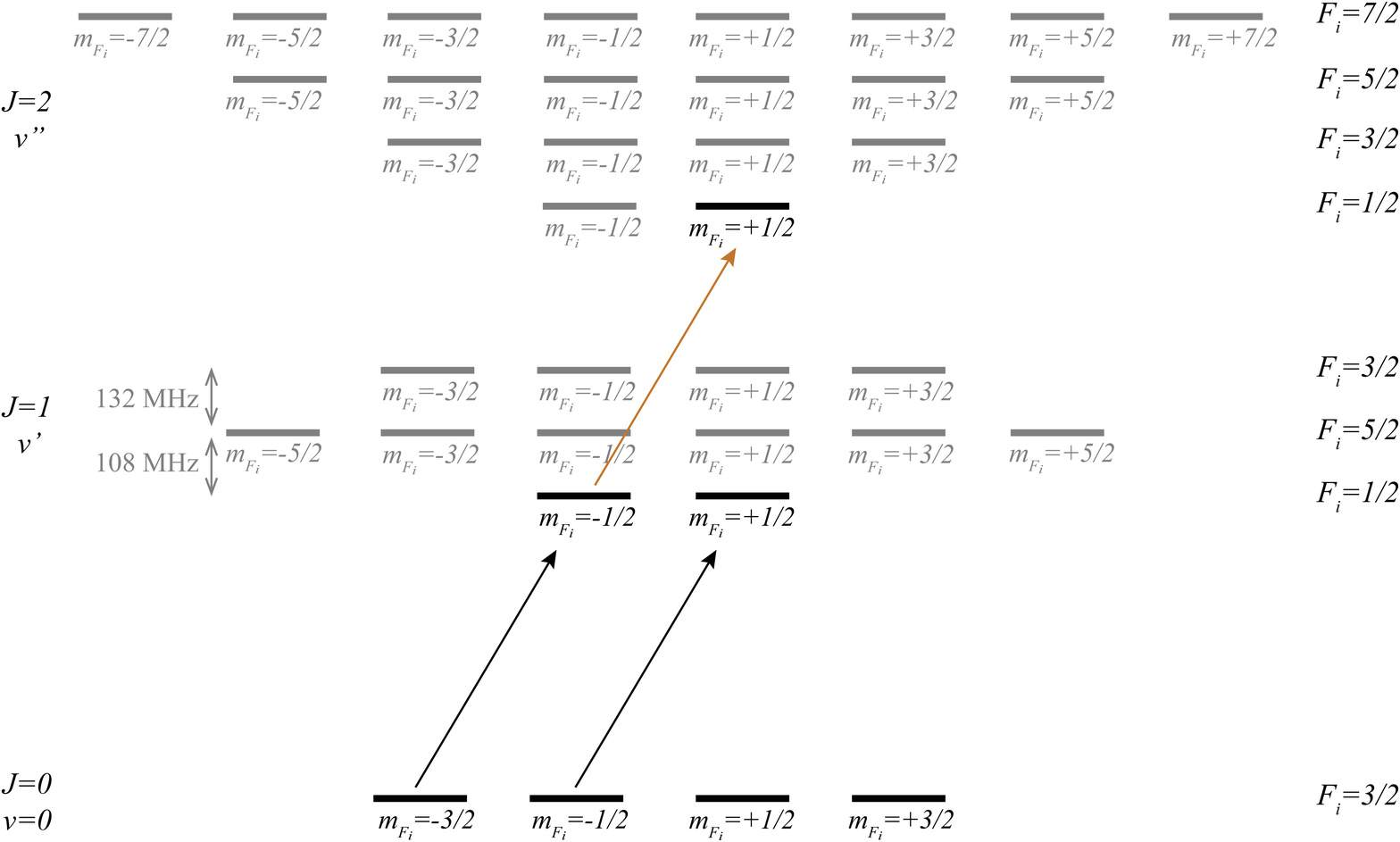}}}\hspace{5pt}
	\subfloat[Polarization beating of $\langle m_{F_i} \rangle$ and $\langle m_{I_2} \rangle$. The hydrogen nucleus is 100\% polarized at $t_0 {=} 4.87$ $\rm{\mu}$s.]{%
		\resizebox*{5.2cm}{!}{\includegraphics{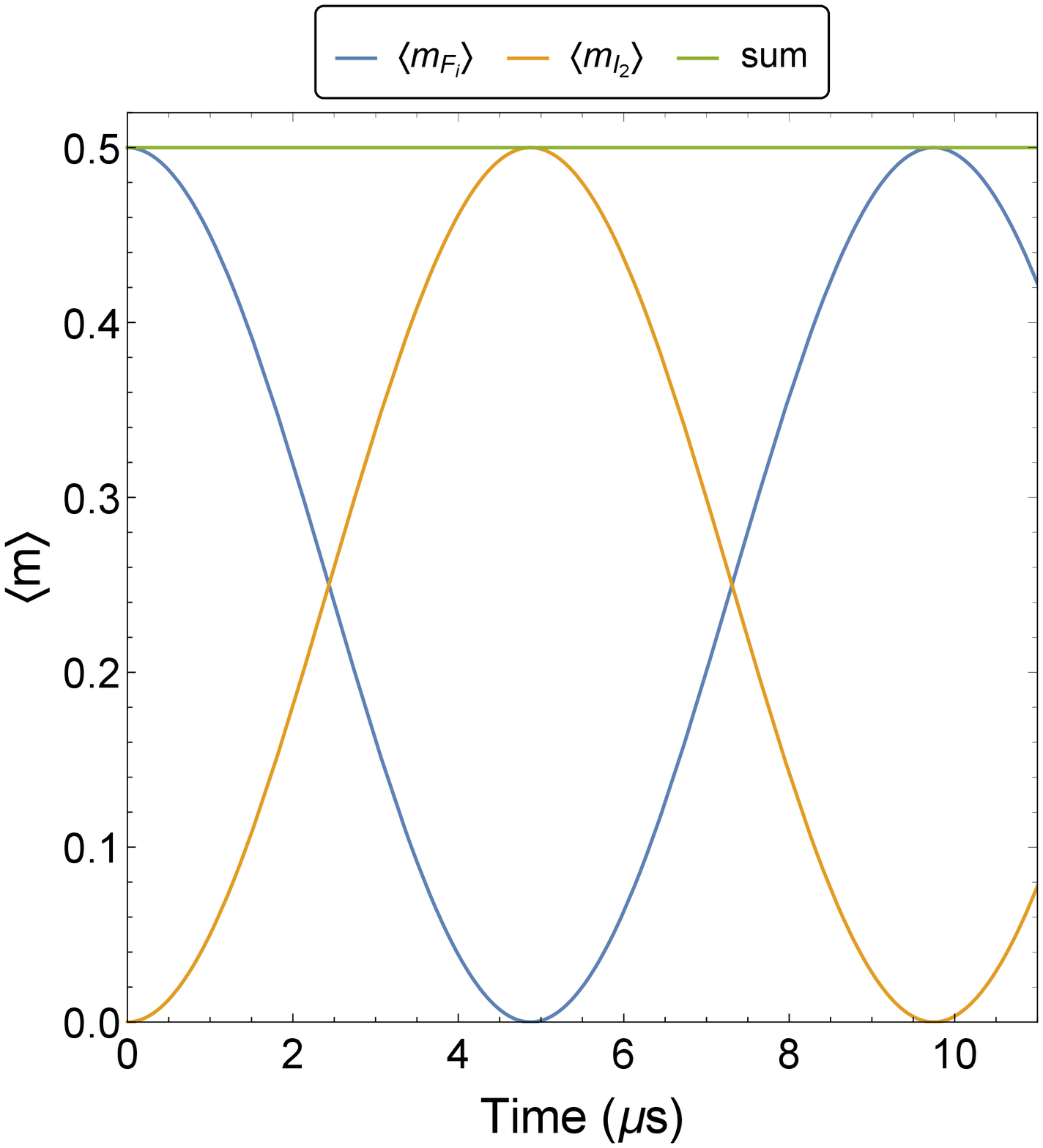}}}
	\caption{IR-excitation steps of H\textsuperscript{79}Br and polarization transfer due to the hyperfine interaction.} \label{fig:HCl}
\end{figure}

We investigate the polarization dynamics of H\textsuperscript{79}Br in partial hyperfine resolution, depicted in Fig.~\ref{fig:HCl}(a). This process can be extrapolated, in general, to molecules with large hyperfine splitting caused by nucleus X, which we do not want to polarize. For the description of our system we adopt the notation: $I_1$ is the bromine nuclear spin, $I_2$ is the proton spin, $J$ is the rotational angular momentum, $F_i = J+ I_1$, and $F= F_i + I_2$. First, the H\textsuperscript{79}Br molecule is excited from the ground state to the $|\nu^{\prime} ,\, J=1,\, F_i = 1/2 \rangle$ ($m_{F_i} =\pm 1/2$) state with right circularly polarized IR light. A second excitation follows to the $|\nu^{\prime\prime} ,\, J=2,\, F_i = 1/2 \rangle$ state with right circularly polarized IR light. Consequently, at $t=0$ the molecular $F_i$ is fully polarized, since only one $m_{F_i}$ substate is populated, i.e. $\langle m_{F_i}\rangle_{t=0} = + 1/2$. However, due to the hyperfine interaction the polarization will be transferred to the proton (at $t_0 {\sim} 4.9$ $\rm \mu s$). At this moment the dissociation of the molecule takes place and the nuclear-spin-polarized hydrogen atoms are produced. We present the theory of polarization dynamics of the H\textsuperscript{79}Br molecule in order to calculate the time when the polarization is transferred to the proton. 

The hyperfine Hamiltonian of heteronuclear diatomic molecules expressed in frequency units can be taken as~\cite{ramseyHCl,wachem}
\begin{equation}\label{eq:halham}
	\begin{split}
	H/h =&-{eq_1 Q_1 \over 2I_1(2I_1 - 1)(2J-1)(2J+3)}\big[{3} (\vec{I}_1 \cdot \vec{J})^2  +{3\over 2}(\vec{I}_1\cdot\vec{J})-\vec{I}_1 ^2\vec{J}^2 \big] 
	-{eq_2 Q_2 \over 2I_2(2I_2 - 1)(2J-1)(2J+3)}\\
	&\times\big[{3} (\vec{I}_2 \cdot \vec{J})^2 +{3\over 2}(\vec{I}_2\cdot\vec{J})
	-\vec{I}_2 ^2\vec{J}^2 \big] 
	+{d_T \over (2J-1)(2J+3)}\big[{3} (\vec{I}_1 \cdot \vec{J})(\vec{I}_2 \cdot \vec{J}) + {3} (\vec{I}_2\cdot\vec{J})(\vec{I}_1 \cdot \vec{J})-2 \vec{I}_1\cdot\vec{I}_2 \vec{J}^2 \big] \\
	&+c_1 \vec{I}_1 \cdot \vec{J} + c_2 \vec{I}_2\cdot \vec{J} +\delta \vec{I}_1 \cdot \vec{I}_2 ,
	\end{split}
\end{equation}
where the first two terms correspond to the interaction of the nuclear quadrupole moments $Q_i$ with the electric field gradients $q_i$ ($Q=0$ for nuclei with spin $1/2$, namely $Q_2 =0$). The third term contains the direct spin-spin interaction and the tensor component of the electron coupled spin-spin interaction. The direct spin-spin interaction constant equals $g_1 g_2 \mu_N ^2 \langle 1/r^{3} \rangle_{\nu,J}$, where $g_1$ and $g_2$ are the nuclear $g$ factors for the two nuclei, $r$ is the internuclear distance, and $\mu_N$ is nuclear magneton. The constants $c_{1,2}$ are the spin-rotation coupling constants and the last term describes the electron-coupled nuclear dipole-dipole interaction.

\begin{figure}
	\centering
	\subfloat[IR-excitation steps of H\textsuperscript{127}I presented in the partial hyperfine resolution. At $t=0$, the $|\nu^{\prime\prime\prime} ,\, J=3,\, F_i = 1/2 \rangle$ state is populated and the molecule is prepared in the $m_{F_i} = + 1/2$ substate. ]{%
		\resizebox*{8.7cm}{!}{\includegraphics{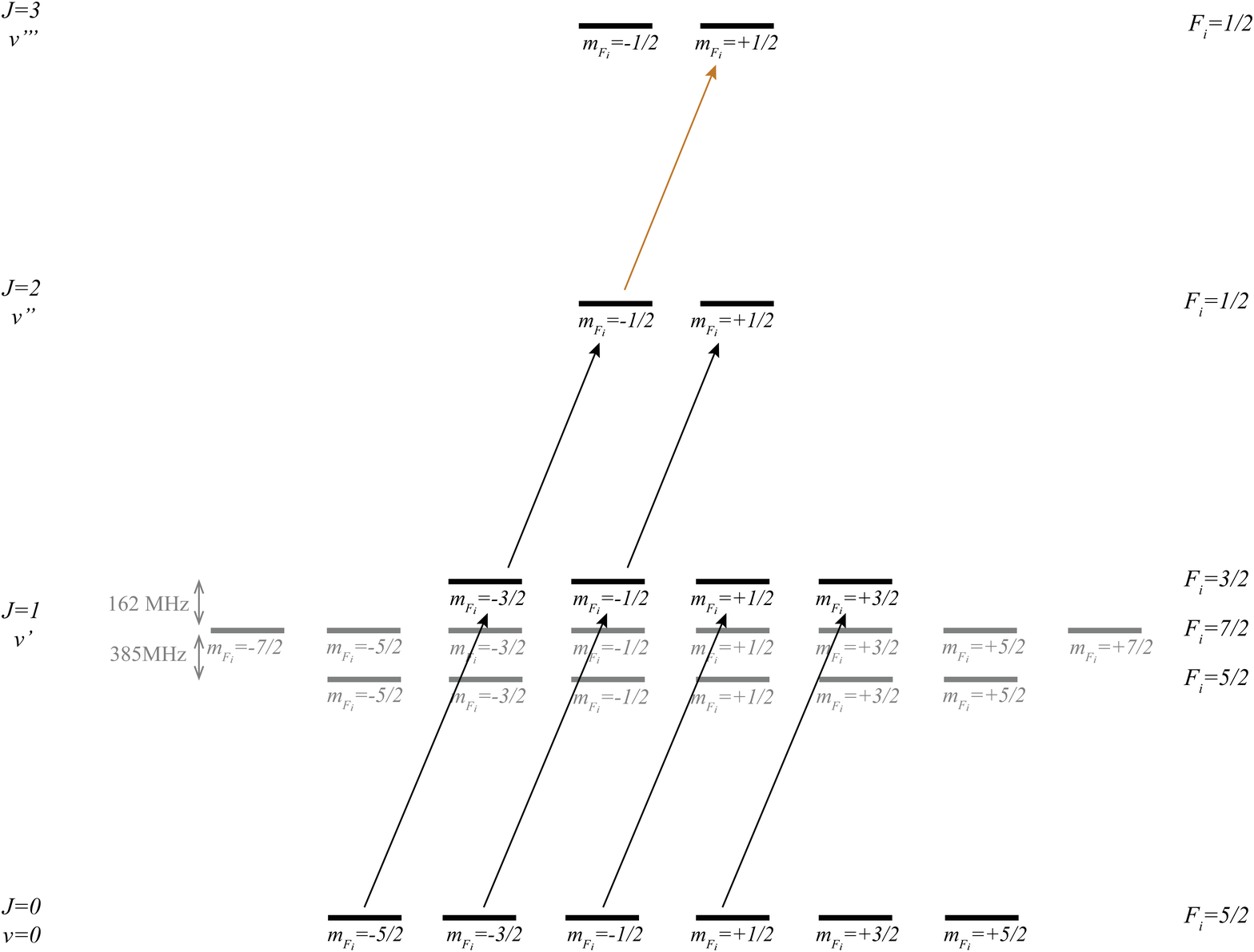}}}\hspace{5pt}
	\subfloat[Polarization beating of $\langle m_{F_i} \rangle$ and $\langle m_{I_2} \rangle$. The hydrogen nucleus is 100\% polarized at $t_0 {=} 3.42$ $\rm{\mu}$s.]{%
		\resizebox*{5.9cm}{!}{\includegraphics{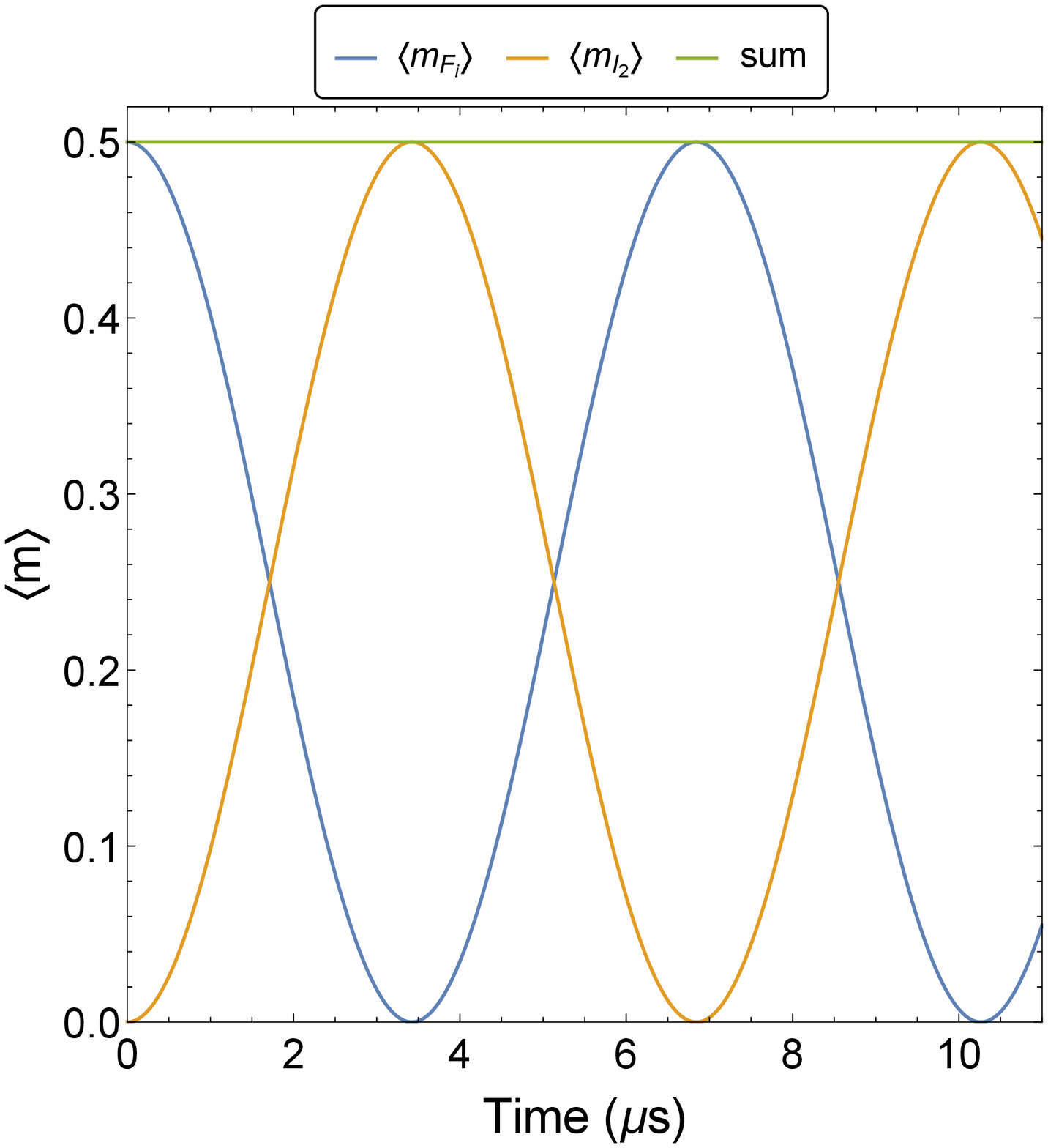}}}
	\caption{IR-excitation steps of H\textsuperscript{127}I and polarization transfer due to the hyperfine interaction.} \label{fig:HI}
\end{figure}

The Hamiltonian is evaluated in the angular momenta projection representation $|m_J,\, m_{I_1},\, m_{I_2}\rangle$ and is diagonalized in order to obtain the time evolved states~\cite{kannis}. However, according to the polarization scheme showed in Fig.~\ref{fig:HCl}(a), the suitable representation is $|F_i,\, m_{F_i};\, m_{I_2}\rangle$ which is connected to the aforementioned representation by a unitary transformation
\begin{equation}\label{eq:clebsg}
|F_i,\, m_{F_i};\, m_{I_2}\rangle = \sum_{m_J, m_{I_1}} \langle J,\, m_J ,\, I_1,\, m_{I_1}|F_i,\, m_{F_i}\rangle |m_J,\, m_{I_1},\, m_{I_2}\rangle.
\end{equation}

\begin{figure}
	\centering
	\subfloat[Absorption spectrum of H\textsuperscript{35}Cl for selected vibrational states ($\nu{=}0,$ $\nu{=}2,$ and $\nu{=}4$). It is shifted to the red by the IR absorption energy.]{%
		\resizebox*{7cm}{!}{\includegraphics{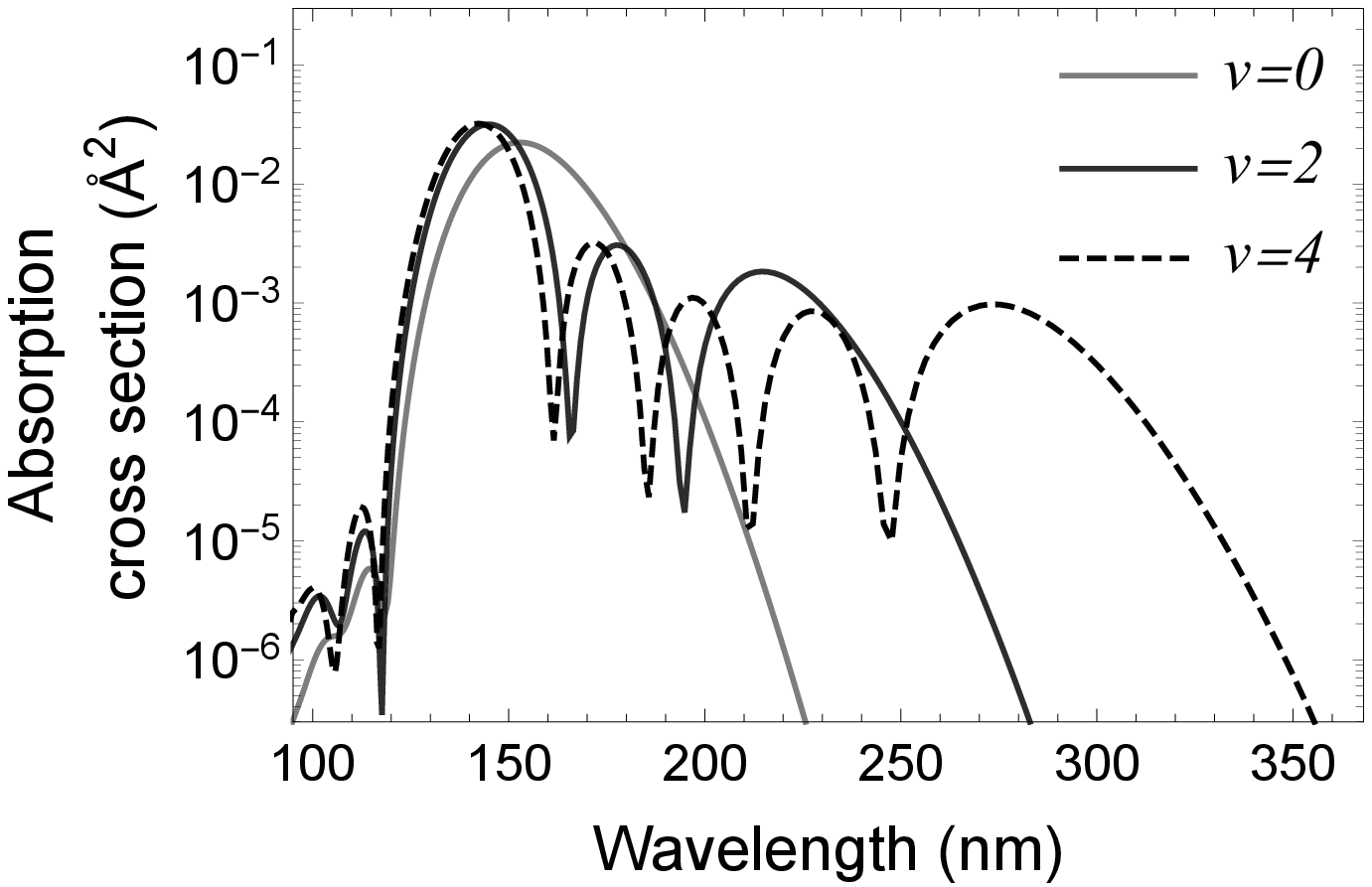}}}\hspace{5pt}
	\subfloat[Absorption spectrum of D\textsuperscript{35}Cl for selected vibrational states ($\nu{=}0,$ $\nu{=}2,$, $\nu{=}4$, $\nu{=}6$, and $\nu{=}8$). It is shifted to the red by the IR absorption energy.]{%
		\resizebox*{7cm}{!}{\includegraphics{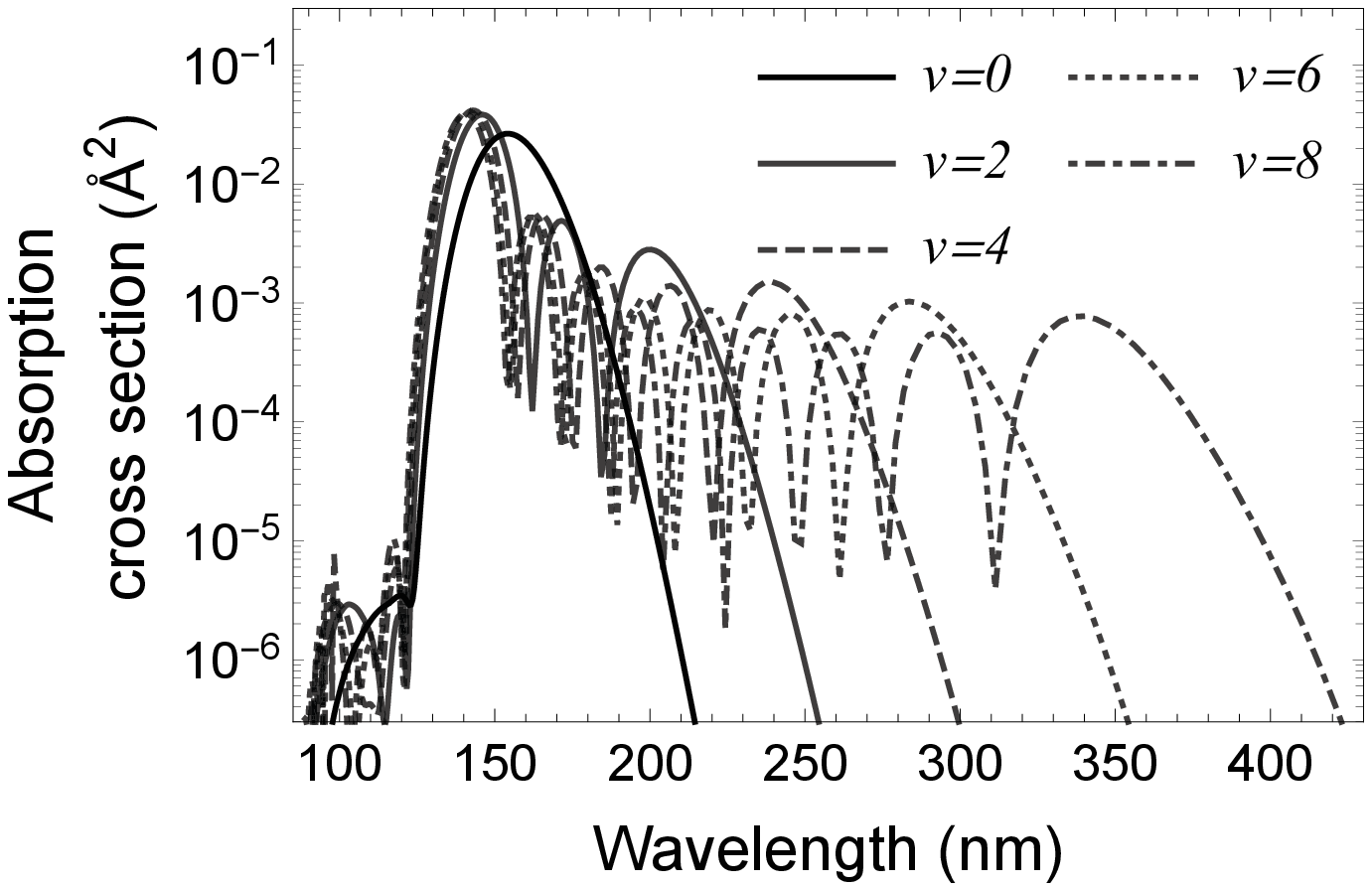}}}\\
	\subfloat[Absorption spectrum of H\textsuperscript{127}I for selected vibrational states ($\nu{=}0,$ $\nu{=}2,$ and $\nu{=}4$). It is shifted to the red by the IR absorption energy.]{%
		\resizebox*{7cm}{!}{\includegraphics{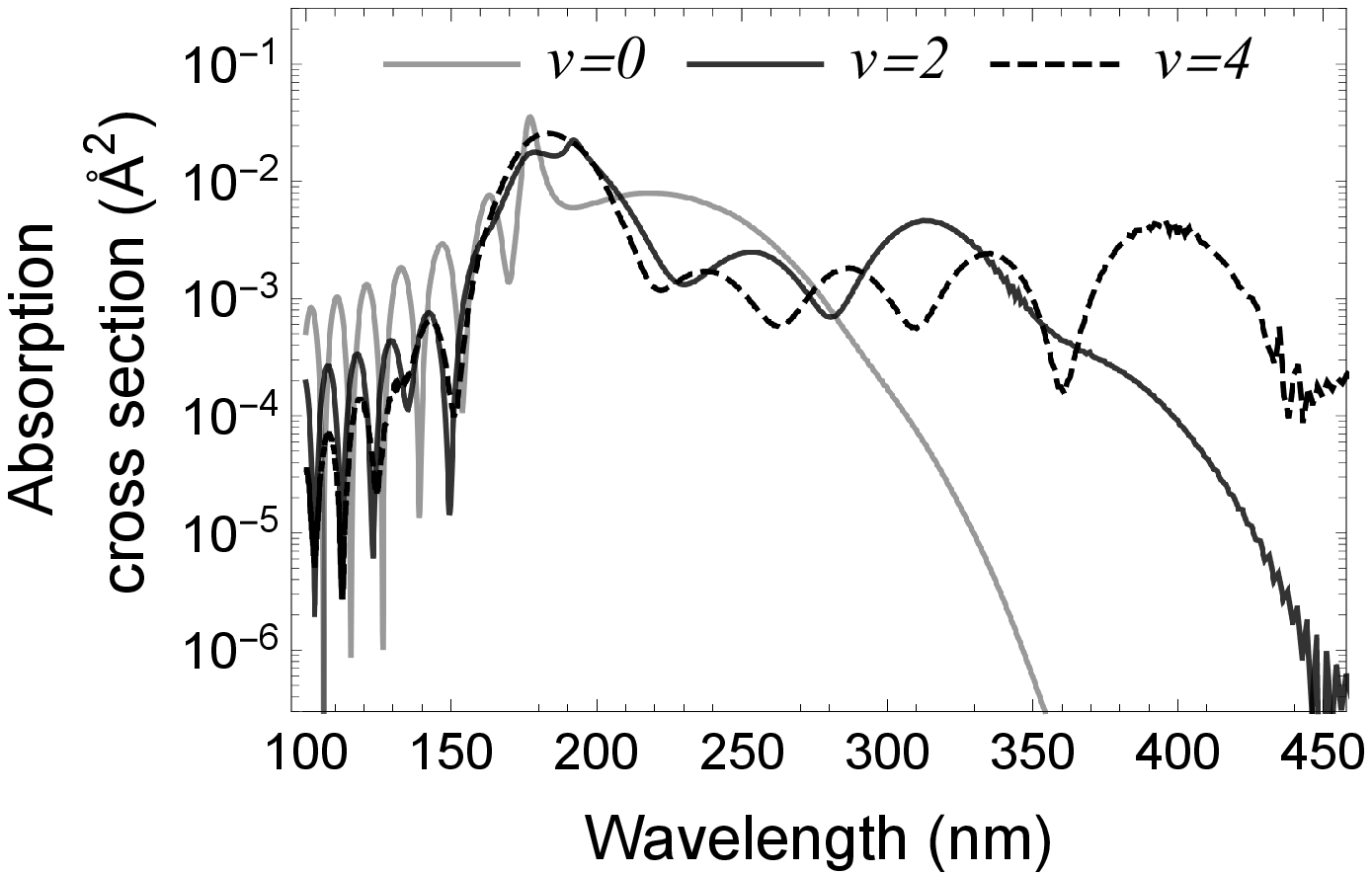}}}\hspace{5pt}
	\subfloat[Absorption spectrum of D\textsuperscript{127}I for selected vibrational states ($\nu{=}0,$ $\nu{=}2,$, $\nu{=}4$, $\nu{=}6$, and $\nu{=}8$). It is shifted to the red by the IR absorption energy.]{%
		\resizebox*{7cm}{!}{\includegraphics{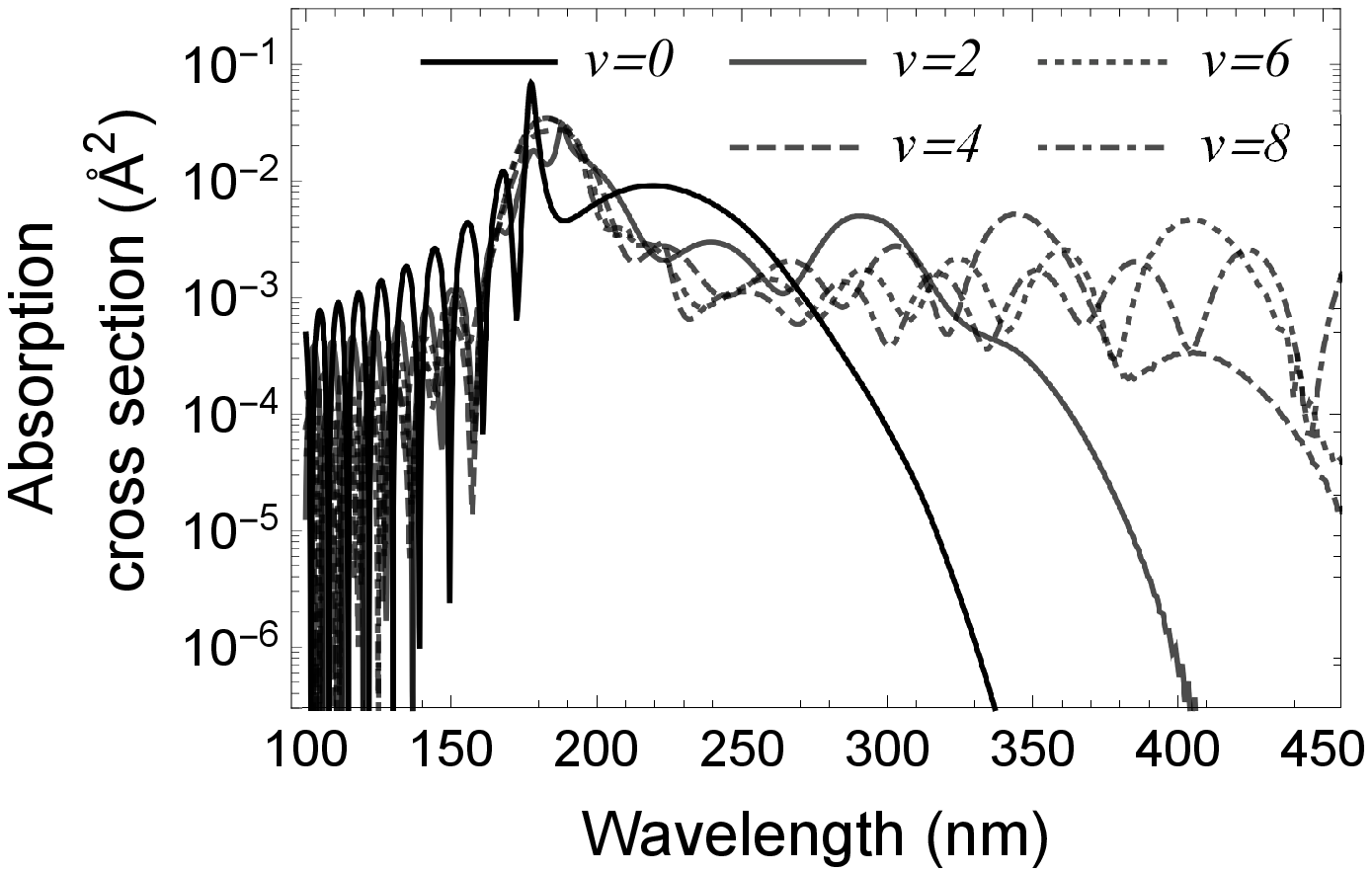}}}
	\caption{Absorption spectrum of H\textsuperscript{35}Cl, D\textsuperscript{35}Cl, H\textsuperscript{127}I, and D\textsuperscript{127}I for selected vibrational states.} \label{fig:cross}
\end{figure}

Based on the above analysis, we plot the polarization beatings of the $|\nu^{\prime\prime} ,\, J=2,\, F_i = 1/2,\, m_{F_i} = +1/2\rangle$ substate (Fig.~\ref{fig:HCl}(b)). The hyperfine constants, used here, correspond to the $\nu = 0,\, J=1$ state of H\textsuperscript{79}Br~\cite{dijk}: $eq_1 Q_1 = 532304.1$ kHz, $c_1 = 290.83$ kHz, $c_2 = -41.27$ kHz, $d_T = 10.03$ kHz. The constant $\delta$ is very small to be measured and is neglected. We expect that the hyperfine constants for higher vibrational states $\nu^{\prime\prime}>0$ do not vary much. Thus, the at $t_0 {\sim} 4.9$ $\rm{\mu}$s the $|\nu^{\prime\prime} ,\, J=2,\, F_i = 1/2 \rangle$ state is going to be fully nuclear-spin-polarized.

Figure~\ref{fig:cross} shows the theoretical predictions for the photodissociation cross section for several vibrational states of HCl, DCl, HI, and DI. The data were obtained by time-propagating several initial vibrational states on the corresponding Potential Energy Curves (PECs) (see~\cite{brownhcldcl,jodoin} and references therein). This methodology builds on a time-dependent approach, where a time-dependent observable is Fourier Transform to yield physical properties that depend on the photon energy. The GridTDSE~\cite{suarez2009,sofsuar} computational codes used in this work incorporates the PECs, the Transition Dipole Moment (TDM) and the electronic couplings between states as input data on the grid of points upon which the wave packet is represented. The cross section is shifted to higher wavelengths for higher lying vibrational levels (the behavior for HBr is similar to the of HCl, shifted about 10-20\% to the red).

\subsection{Production of 60\%-70\%  D polarization from DBr}

\begin{figure}
\centering
\includegraphics*[width=.9\textwidth]{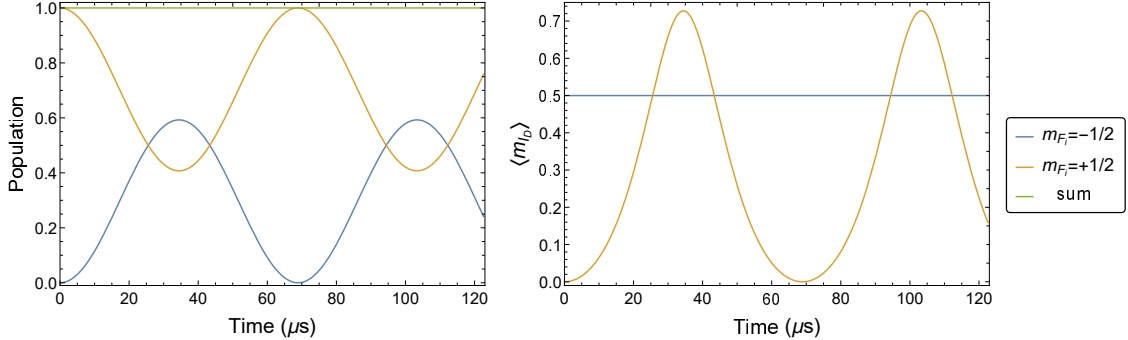}
\caption{D\textsuperscript{79}Br: Population and $\langle m_{I_D}\rangle$ of $m_{F_i}$ states. The $m_{F_i} = + 1/2$ is initially populated and at $t_0 = 34.4\, {\rm \mu s}$ its polarization is maximized (72.7\%).} \label{fig:DCl60pc}
\end{figure}

The IR-excitation steps depicted in Fig.~\ref{fig:HCl}(a) can be applied for the production of ${\gtrsim}60\%$ polarized D. At $t=0$, the $|\nu^{\prime\prime} ,\, J=2,\, F_i = 1/2 ,\, m_{F_i} = + 1/2 \rangle$ substate is populated and the D nuclear spin is unpolarized. The hyperfine dynamics are described by Eq.~\ref{eq:halham} and the hyperfine constants for the $\nu=0,\, J=1$ state are~\cite{phddijk}: $eq_1 Q_1 = 530631.5$ kHz, $eq_2 Q_2 = 146.9$ kHz, $c_1 = 145.82$ kHz, $c_2 = -3.25$ kHz, and $d_T = 1.59$ kHz. Figure~\ref{fig:DCl60pc} shows the polarization dynamics of the $m_{F_i}$ states. The nuclear polarization is maximized (59.3\%) at $t_0 = 34.4\, {\rm \mu s}$. The population of $m_{F_i} = - 1/2$ and $m_{F_i} = + 1/2$ is 59.26\% and 40.74\%, respectively. At this moment we can either photodissociate the molecules or apply an IR excitation with left circularly polarized light to $|\nu^{\prime\prime\prime} ,\, J=1,\, F_i = 1/2 ,\, m_{F_i} = - 1/2 \rangle$, so that only the molecules in the $m_{F_i} = +1/2$ state will follow. Next, one more IR transition is required to a higher vibrational level with $J=0$ to avoid hyperfine depolarization. In this case the D polarization is 72.7\%.  

\section{Production of 100\%  D polarization from DBr}\label{dcl100pc}

\begin{figure}
	\centering
	\subfloat[IR-excitation steps of D\textsuperscript{79}Br for achieving 100\% D polarization, in the partial hyperfine resolution.]{%
		\resizebox*{8.6cm}{!}{\includegraphics{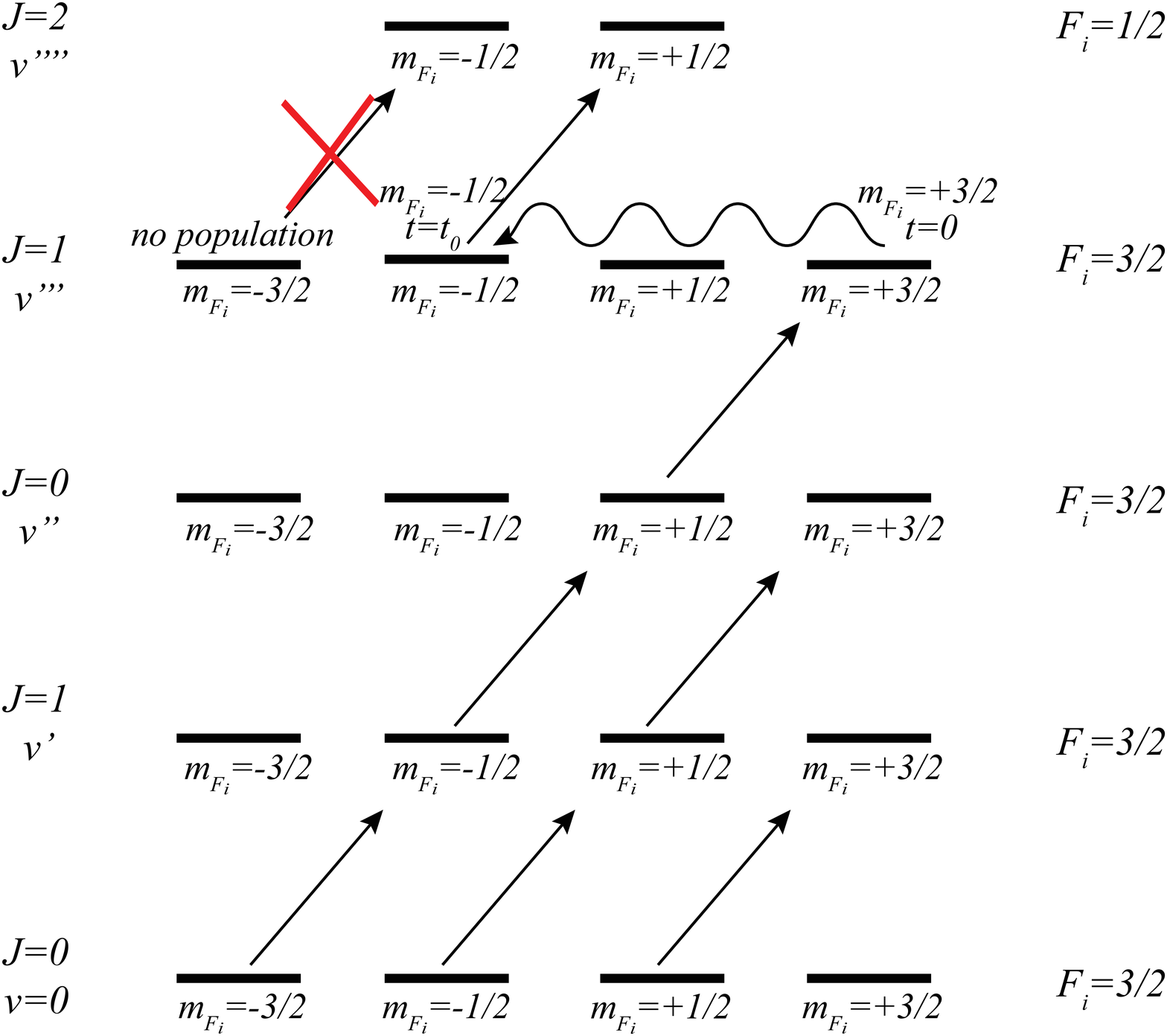}}}\hspace{5pt}
	\subfloat[Population and $\langle m_{I_D} \rangle$ of $m_{F_i}$ states as function of time. The $m_{F_i} {=} -1/2$ state has 100\% polarized D and reaches ${\sim} 25.8\%$ of population at $t_0 {=} 26.6$ $\rm{\mu s}$.]{%
		\resizebox*{5.4cm}{!}{\includegraphics{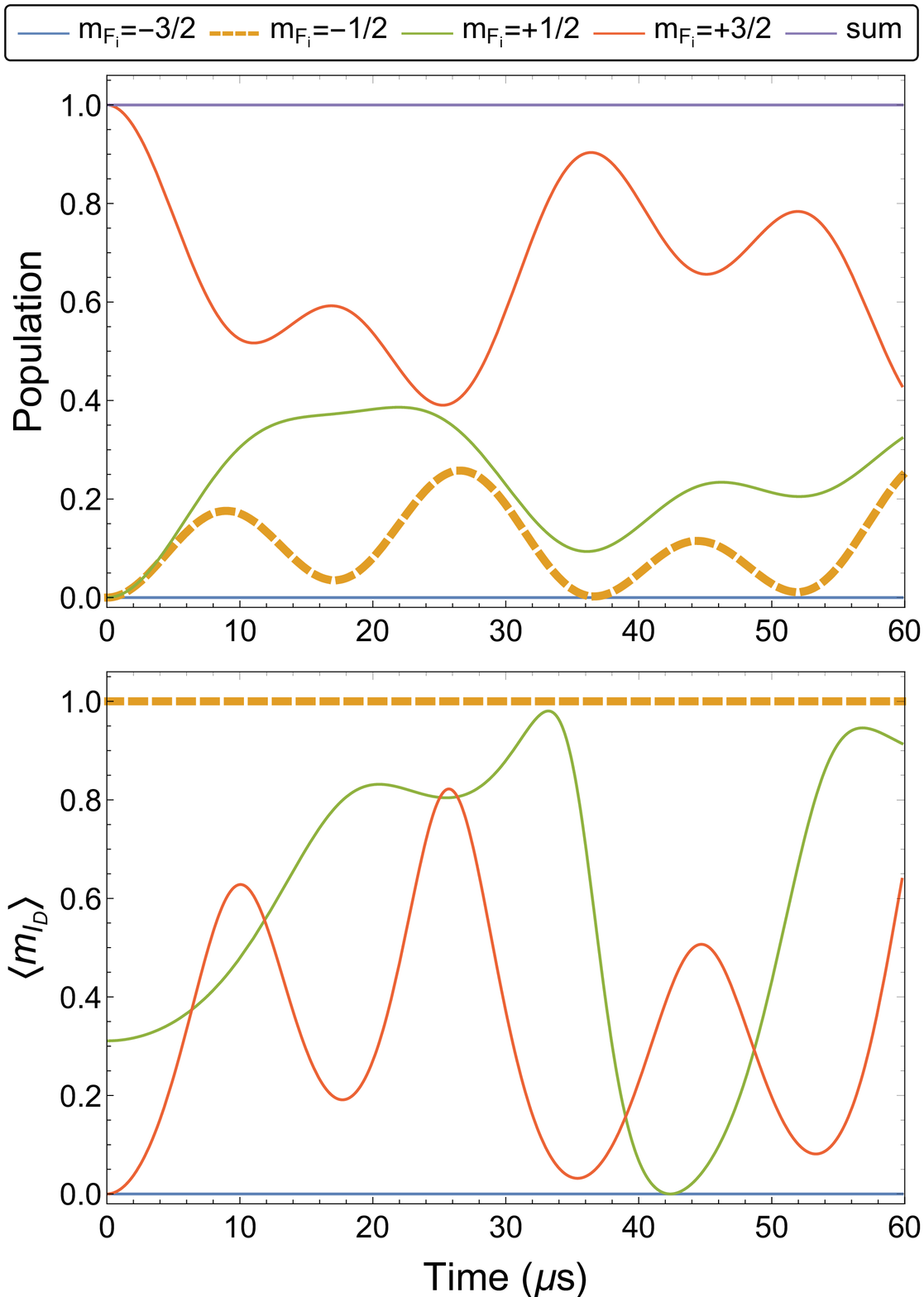}}}
	\caption{IR-excitation steps of D\textsuperscript{79}Br and polarization transfer due to the hyperfine interaction.} \label{fig:DCl}
\end{figure}

The production of 100\% D polarization is feasible via a 4-photon IR-excitation and photodissociation of D\textsuperscript{79}Br (Fig.~\ref{fig:DCl}(a)). After the  absorption of 3 right circularly polarized photons the molecule is excited (from the ground state) to the $|\nu^{\prime\prime\prime} ,\, J=1,\, F_i = 3/2 , \, m_{F_i}=+ 3/2  \rangle$ substate. Then, we leave the system to evolve freely, transferring polarization to D, until the population of $|\nu^{\prime\prime\prime} ,\, J=1,\, F_i = 3/2 , \, m_{F_i}=- 1/2  \rangle$ is maximized ($t=t_0$). The $|\nu^{\prime\prime\prime} ,\, J=1,\, F_i = 3/2 , \, m_{F_i}=- 3/2  \rangle$ state is not populated (due to angular momentum projection conservation). Thus, the absorption of a right circularly polarized photon, at $t=t_0$, excites the molecule to $|\nu^{\prime\prime\prime\prime} ,\, J=2,\, F_i = 1/2 , \, m_{F_i}=+ 1/2  \rangle$. Two more transitions are required directly afterwards to reach a higher vibrational level with $J=0$ and maintain nuclear polarization.

The hyperfine dynamics are described by Eq.~\ref{eq:halham} and the hyperfine constants for the $\nu=0,\, J=1$ state can be found in~\cite{phddijk}. In the same manner with H\textsuperscript{79}Br, we calculate the population and nuclear-spin-polarization beatings (Fig.~\ref{fig:DCl}(b)). The $\langle m_{I_D} \rangle$ of each state has been calculated as the ratio of the total $\langle m_{I_D} \rangle$ of the molecule to the population of each $m_J$ state (when is nonzero) and it indicates the amount of nuclear polarization for each state. Figure~\ref{fig:DCl}(b) shows that the population of $m_{F_i} = -1/2$ reaches ${\sim} 25.8\%$, at $t_0 = 26.6$ $\rm{\mu s}$. 

\section{Hyperfine Dynamics of Ammonia}\label{ammonia}

\begin{figure}
	\centering
	\subfloat[IR-excitation steps for achieving 100\% or 87.6\% (gray) H polarization.]{%
		\resizebox*{8.6cm}{!}{\includegraphics{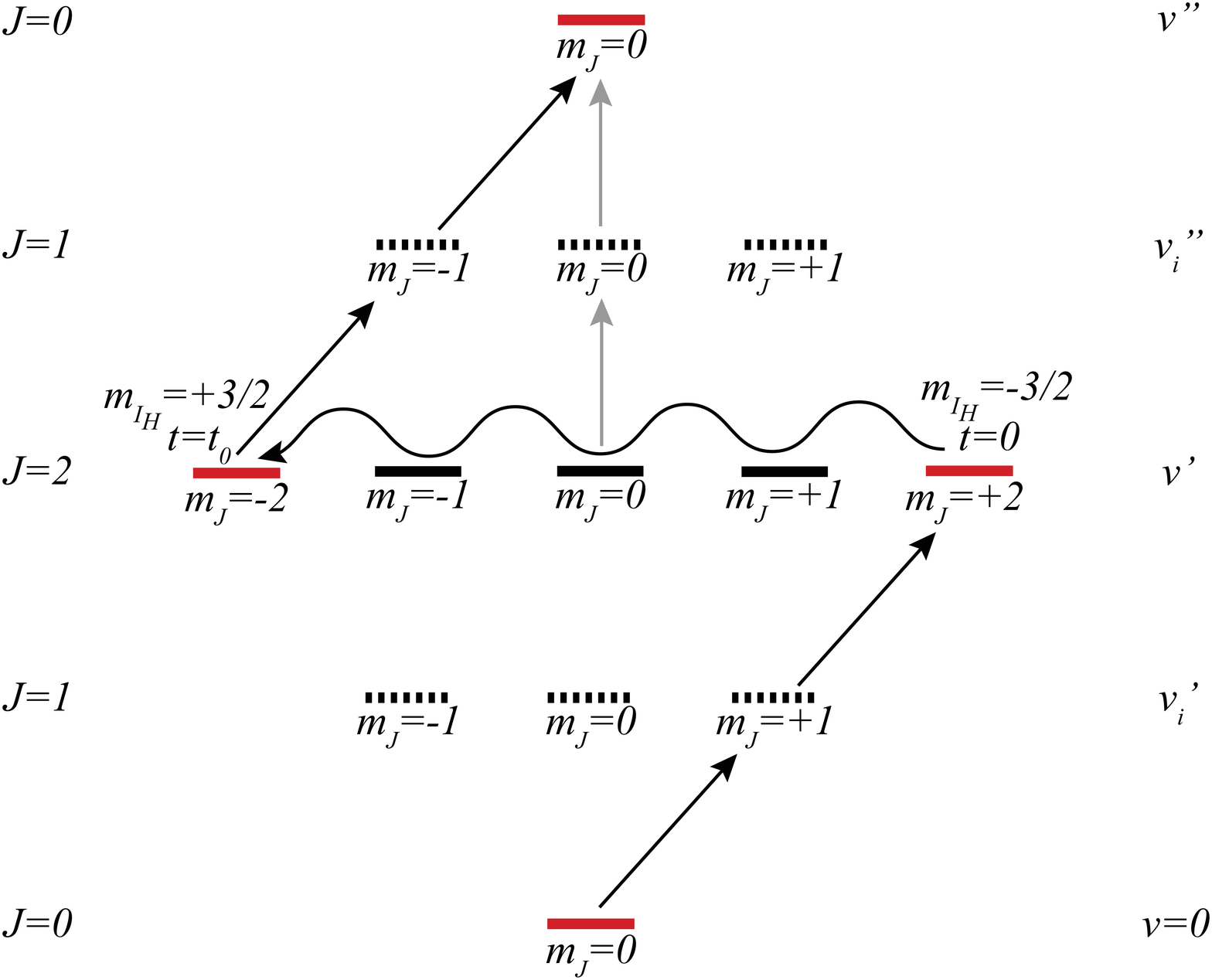}}}\hspace{5pt}
	\subfloat[Population and $\langle m_{I_H} \rangle$ of $m_{J}$ states as function of time. The $m_{J}{=}-2$ state has 100\% polarized H and reaches $3.5\%,\, 4.9\%,\, 10.9\%$ of population at $t_0{=}18.7,\, 34.1,$ and $68.9$ $\rm{\mu s}$, respectively. ]{%
		\resizebox*{5.4cm}{!}{\includegraphics{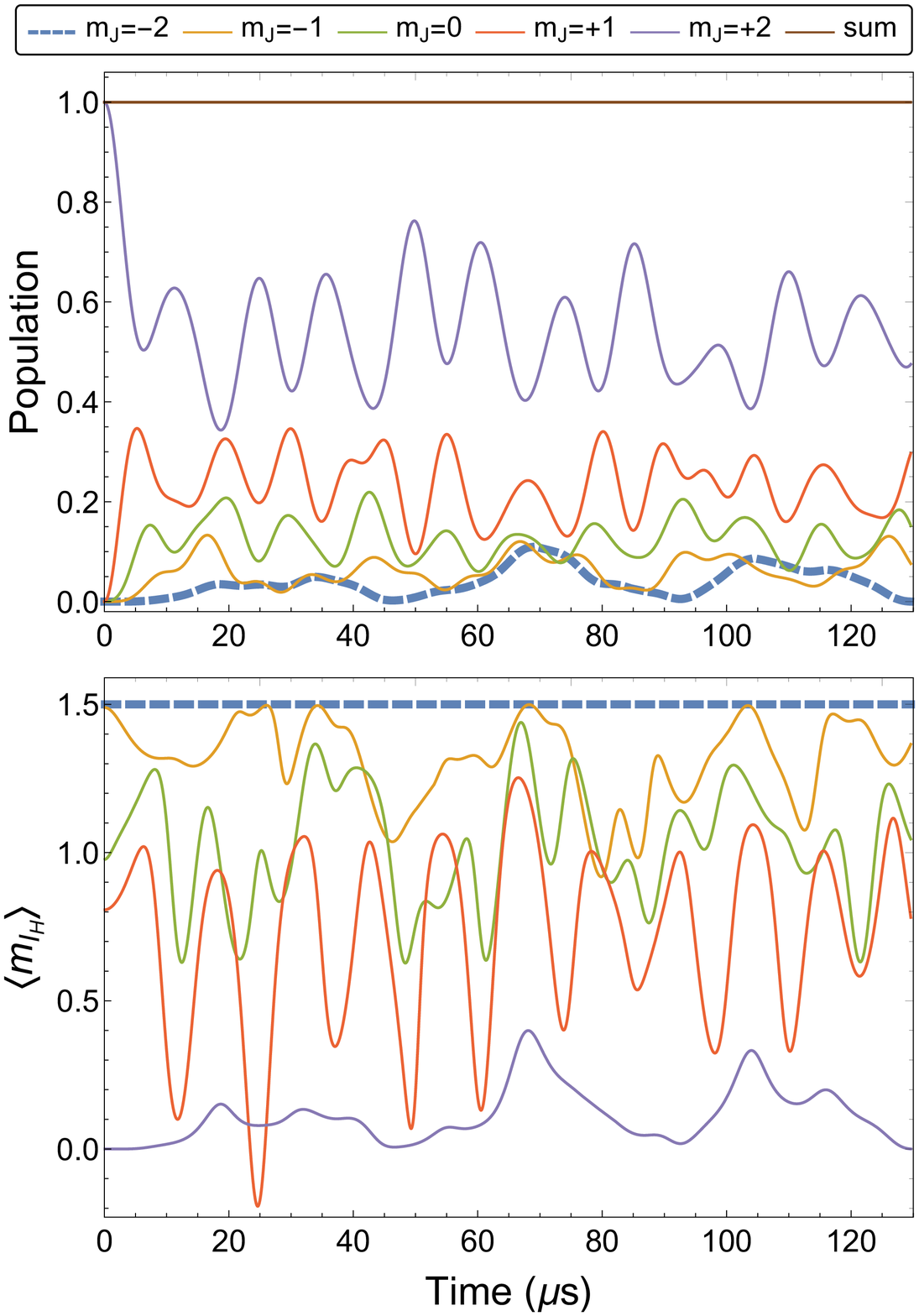}}}
	\caption{IR-excitation steps of \textsuperscript{15}NH\textsubscript{3} and polarization transfer due to the hyperfine interaction.} \label{fig:NH3}
\end{figure}

Ammonia, NH\textsubscript{3}, is a trigonal pyramid with three identical N-H bonds. It exists in two (ortho and para) forms which correspond to different values of the total nuclear spin of the hydrogen atoms ($I_H =3/2$ and $1/2$ respectively). The requirement that the total wavefunction of ammonia should be antisymmetric under interchange of hydrogen nuclei implies that the value of $K$ for ortho-ammonia is a multiple of three whereas para-ammonia has values of $K=3n\pm 1$. Figure~\ref{fig:NH3}(a) shows the IR-excitation process for achieving 100\%  (via $m_J = -2$) or 87.6\% (via $m_J =0$) H polarization.

The effective hyperfine Hamiltonian of \textsuperscript{15}NH\textsubscript{3} has been studied by Hougen and Kukolich~\cite{hougen,kukolich68}:

\begin{equation}\label{eq:ammonia}
\begin{split}
H_{eff}/h=& R \vec{I}_N \cdot \vec{J} + S \vec{I}_H \cdot \vec{J} +\frac{2T}{(2J-1)(2J+3)} \Bigg[\frac{3}{2} (\vec{I}_N \cdot \vec{J}) (\vec{I}_H \cdot \vec{J}) +\frac{3}{2}(\vec{I}_H \cdot \vec{J})(\vec{I}_N \cdot \vec{I}_H)-(\vec{I}_N\cdot \vec{I}_H)\vec{J}^2\Bigg] \\
& +\frac{2U}{(2J-1)(2J+3)}\Big[3(\vec{I}_H \cdot \vec{J})^2 +\frac{3}{2}(\vec{I}_H\cdot \vec{J})-\vec{I}_H ^2 \vec{J}^2\Big], 
\end{split}
\end{equation}
where $\vec{J},$ $\vec{I}_N$ and $\vec{I}_H$ correspond to the total rotational angular momentum, the spin of the nitrogen nucleus, and the total nuclear spin of the three protons (in units of $\hbar$), respectively. The parameters describe the interactions: $R$: N spin-rotation coupling, $S$: H spin-rotation coupling, $T$: N-H spin-spin interaction, and $U$: H-H spin-spin interaction. The formulas are defined in~\cite{hougen,kukolich67,kukolich68} using the parameters introduced by~\cite{gordon,gunther}.

We are interested in the polarization dynamics of the ortho $J=2$ state, i.e. $J=2$, $K=0$. However, this state has not been studied experimentally. Nonetheless, we can deduce the values of the hyperfine parameters from the existing data of the $J-K=1-1,\, 2-2,\, 3-3,\,  {\rm and}\,\, 4-4$ states~\cite{kukolich68,hougen}. Thus, we substitute $R{\sim}-9.68$ kHz, $S{\sim}-17.8$ kHz, $T{\sim}3.33$ kHz, and $U{\sim}-6.9$ kHz into Eq.~\ref{eq:ammonia}. The aforementioned constants can be verified by the data of the $J=1$, $K=0$ state of \textsuperscript{14}NH\textsubscript{3}~\cite{marshall,cazzoli} and the approximate $^{15}NH_3/^{14}NH_3$ ratios~\cite{hougen} for pairs of constants characterized by the same rotational quantum numbers $J$ and $K$.

Figure~\ref{fig:NH3}(b) shows that 100\% H polarization is possible, but with population less than 11\%, or 87.6\% ($\langle m_{I_H} \rangle_{t_0} = 1.324$) with population 13.3\% at $t_0 = 16.5$ $\rm{\mu s}$.

\subsection{Hyperfine Dynamics of Deuterated Ammonia}

The hyperfine structure of mono-deuterated ammonia, NH\textsubscript{2}D, has been studied in the 60s~\cite{thaddeus,kukolichstephen}. However, these studies were focused on higher-$J$ transitions. Recently, Melosso et al.~\cite{melosso} investigated the $J_{K_{-1} , K_1} = 1_{1,1}-1_{0,1}$ and $1_{0,1}-0_{0,0}$ transitions and reported the required hyperfine constants to construct the hyperfine Hamiltonian of the $1_{1,1}$ state.

NH\textsubscript{2}D exists in two species, ortho and para, depending on the total hydrogen nuclear spin. According to~\cite{weiss}, the lower inversion states with $K_{-1}={\rm odd}$ are ortho and with $K_{-1}={\rm even}$ are para species. In a similar way, the upper inversion states with $K_{-1}={\rm even}$ are ortho and with $K_{-1}={\rm odd}$ are para species.

The hyperfine Hamiltonian is written in the form~\cite{thaddeus}
\begin{equation}\label{eq:nh2d}
\begin{split}
H/h=& \frac{(eq_J Q)_N}{2 I_N (2 I_N -1)J(2J-1)}\Big[3 (\vec{I}_N \cdot \vec{J})^2 +\frac{3}{2}(\vec{I}_N \cdot \vec{J}) - \vec{I}_N ^2 \vec{J}^2\Big]\\
&\frac{(eq_J Q)_D}{2 I_D (2 I_D -1)J(2J-1)}\Big[3 (\vec{I}_D \cdot \vec{J})^2 +\frac{3}{2}(\vec{I}_D \cdot \vec{J}) - \vec{I}_D ^2 \vec{J}^2\Big]\\
&+ C_N (\vec{I}_N \cdot \vec{J}) + C_D (\vec{I}_D \cdot \vec{J}) + C_H (\vec{I}_H \cdot \vec{J})\\
&+\frac{d^{ND} _J}{J(2J-1)}\Bigg[\frac{3}{2} (\vec{I}_N \cdot \vec{J}) (\vec{I}_D \cdot \vec{J}) +\frac{3}{2}(\vec{I}_D \cdot \vec{J})(\vec{I}_N \cdot \vec{I}_D)-(\vec{I}_N\cdot \vec{I}_D)\vec{J}^2\Bigg]\\
&+\frac{d^{NH} _J}{J(2J-1)}\Bigg[\frac{3}{2} (\vec{I}_N \cdot \vec{J}) (\vec{I}_H \cdot \vec{J}) +\frac{3}{2}(\vec{I}_H \cdot \vec{J})(\vec{I}_N \cdot \vec{J})-(\vec{I}_N\cdot \vec{I}_H)\vec{J}^2\Bigg]\\
&+\frac{d^{DH} _J}{J(2J-1)}\Bigg[\frac{3}{2} (\vec{I}_D \cdot \vec{J}) (\vec{I}_H \cdot \vec{J})+\frac{3}{2} (\vec{I}_H \cdot \vec{J})(\vec{I}_D \cdot \vec{J})-(\vec{I}_D\cdot \vec{I}_H)\vec{J}^2\Bigg]\\
&+\frac{d^{HH} _J}{2 I_H (2 I_H -1) J(2J-1)}\Big[3 (\vec{I}_H \cdot \vec{J})^2 + \frac{3}{2}(\vec{I}_H \cdot \vec{J})-\vec{I}_H ^2\vec{J}^2\Big],
\end{split}
\end{equation}
where the first two terms are the N and D quadrupole interactions and the next three terms are the N, D, and H spin-rotation interactions. The rest correspond to the spin-spin interactions of the four nuclei (in pairs). The hyperfine constants are defined as: $(eq_J Q)_K = \frac{2}{(J+1)(2J+3)}\sum_{g} \chi^K _{gg} \langle J_g ^2\rangle$, $C_K = \sum_{g} \frac{C^K _{gg} \langle J_g ^2\rangle}{J (J+1)}$, and $d^{KL} _J =-\frac{2 \mu_N ^2 g_K g_L}{(J+1)(2J+3)}\sum_{g} \frac{3 r_{KL, gg} ^2 \langle J_g ^2\rangle - r_{KL} ^2 \langle J_g ^2\rangle}{r_{KL} ^5}$. The index $g$ represents the principal inertial axes of the molecule $a$, $b$, and $c$. The $\langle J_{g} ^2 \rangle$ are the average values of the square of the components of $J$ along the principal axes, $\chi_{gg} = eQ\frac{\partial^2 V}{\partial g^2}$ is the common nuclear quadrupole constant, $r_{_{KL}}$ is the distance between the nuclei $K$ and $L$ and $r_{_{KL, gg}}$ is the magnitude of the projection of $\vec{r}_{_{KL}}$ (vector joining the two nuclei) on the principal axis $g$. Finally, $C^K _{gg}$ are the spin-rotation constants along the principal inertial axes.

\begin{figure}
	\centering
	\subfloat[$1_{1,1}$-lower inversion state of \textsuperscript{15}NH\textsubscript{2}D: Population and $\langle m_{I_D} \rangle$ of $m_{J}$ states as function of time. The $m_{J} = -1$ state reaches ${\sim}59\%$ D polarization at $t_0 = 22.8$ $\rm{\mu s}$.]{%
		\resizebox*{7cm}{!}{\includegraphics{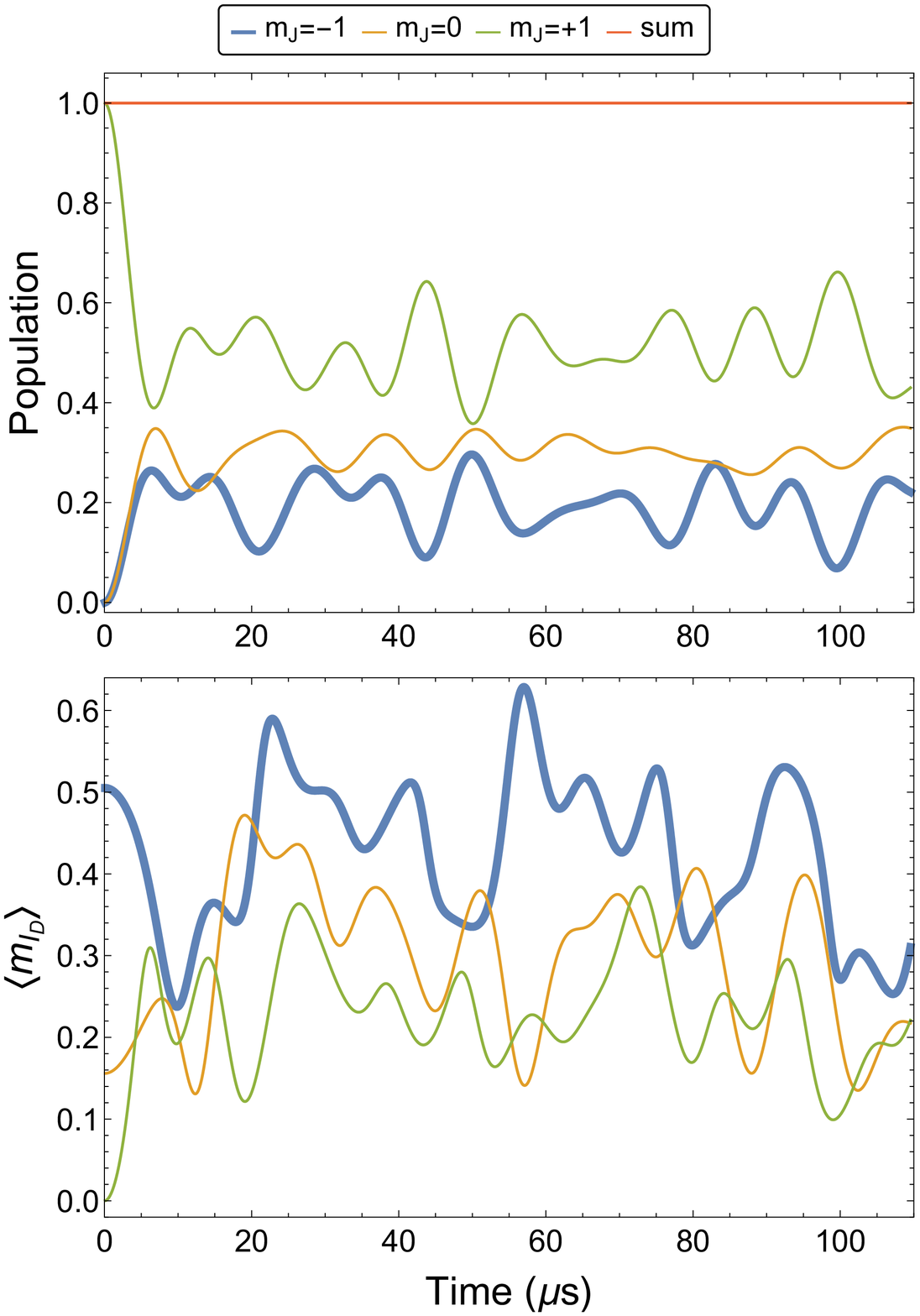}}}\hspace{5pt}
	\subfloat[$1_{1,1}$-upper inversion state of \textsuperscript{15}NH\textsubscript{2}D: Population and $\langle m_{I_D} \rangle$ of $m_{J}$ states as function of time. The population of the $m_{J} {=}-1$ state reaches $19.78\%$ of population at $t_0{=}7.3$ $\rm{\mu s}$. The D polarization by that time is 99.1\%.]{%
		\resizebox*{7cm}{!}{\includegraphics{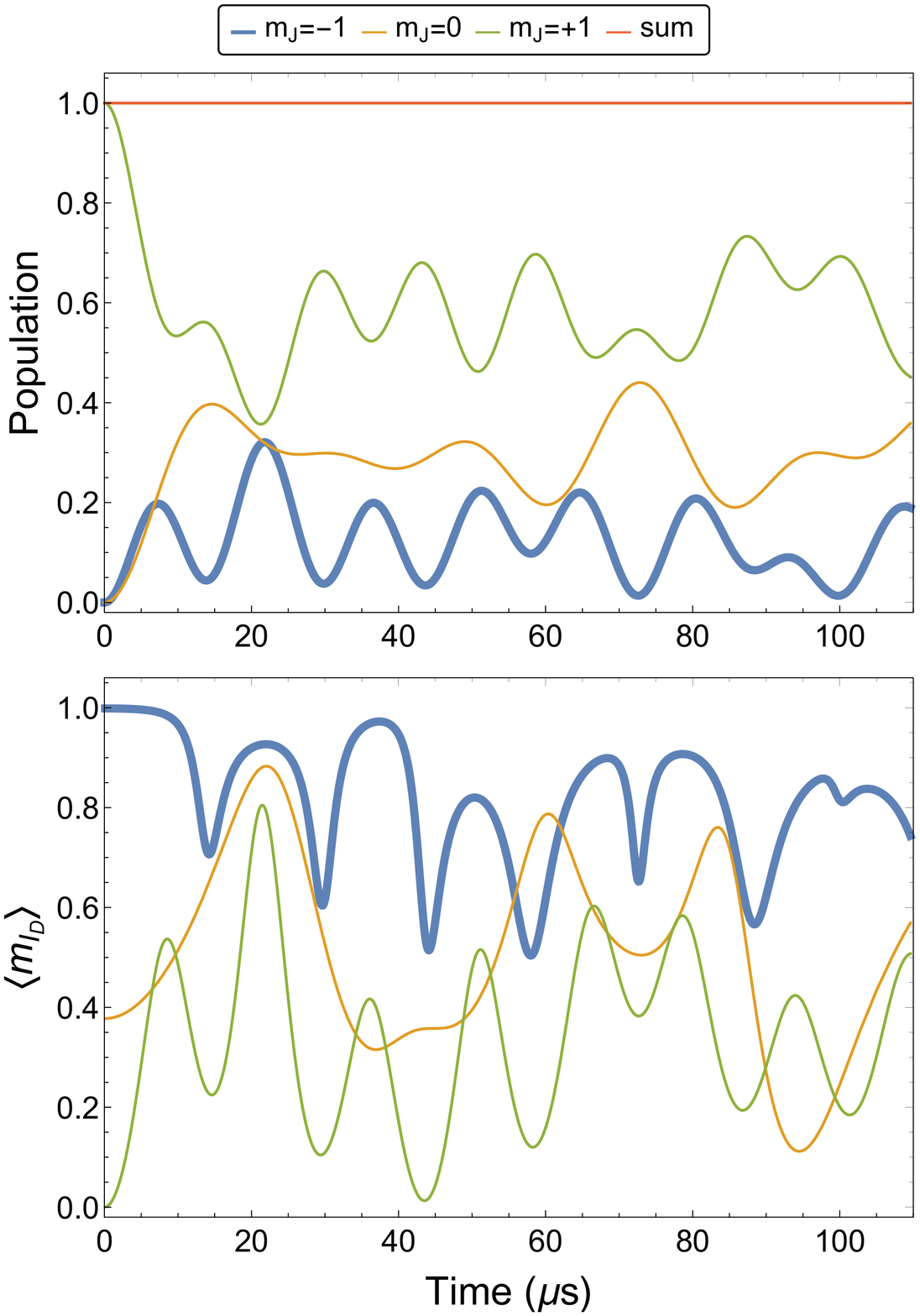}}}
	\caption{Population and $\langle m_{I_D} \rangle$ of $m_{J}$ states as function of time for ortho and para species.} \label{fig:NH2D}
\end{figure}

The hyperfine Hamiltonian of the $1_{1,1}$ state of \textsuperscript{14}NH\textsubscript{2}D can be obtained from the existing (measured and calculated) data of~\cite{melosso}. Since, the \textsuperscript{14}N nucleus has $I_N = 1$ and strong quadrupole interaction, a large portion of the rotational polarization will be transferred to it. Therefore, a more suitable molecule for the production of polarized D is the \textsuperscript{15}NH\textsubscript{2}D. The corresponding Hamiltonian can be evaluated from the \textsuperscript{14}NH\textsubscript{2}D Hamiltonian, by multiplying the constants with the following ratios: $\frac{(eq_J Q)_{_{^{^{15}}N}}}{(eq_J Q)_{_{^{^{14}}N}}}=0$, $\frac{C_{_{^{^{15}}N}}}{C_{_{^{^{15}}N}}} = \frac{d^{^{15}ND} _J}{d^{^{14}ND} _J} = \frac{d^{^{15}NH} _J}{d^{^{14}NH} _J} {\approx} \frac{g_{_{^{15}N}}}{g_{_{^{14}N}}}{\sim}-1.4$. The rest of the constants are invariant to nitrogen isotopic substitution, since they do not involve the nitrogen nucleus. According to Hougen~\cite{hougen} the small deviations from these ratios (due to the different intramolecular distances for the two molecules) are of the order of 1\% for the ammonia molecule. Similar deviations are expected in the NH\textsubscript{2}D molecule.

Consequently, substituting the (measured and computed) values of the hyperfine constants we investigate the polarization dynamics of the \textsuperscript{15}NH\textsubscript{2}D molecule excited with a right circularly polarized photon in the $1_{1,1}$, $m_J = +1$ substate.

Figure~\ref{fig:NH2D}(a) shows the polarization dynamics of the lower inversion $1_{1,1}$ state. Since, it is an ortho state, i.e. $I_H =1$, the D cannot get fully polarized, because a portion of the rotational polarization will also be transferred to the H. Thus, the $m_{J} = -1$ state reaches ${\sim}59\%$ D polarization at $t_0 = 22.8$ $\rm{\mu s}$. The population of this state is 12.58\% at $t_0 = 22.8$ $\rm{\mu s}$. On the other hand, the upper inversion $1_{1,1}$ state belongs to the para species ($I_H =0$). This allows higher D polarization (see Fig.~\ref{fig:NH2D}(b)). At $t_0 = 7.3$ $\rm{\mu s}$ the D polarization and population are 99.1\% and 19.78\%, respectively. At $t_0 = 21.8$ $\rm{\mu s}$ the D polarization and population are 92.68\% and ${\sim}32.1\%$, respectively.

\section{Conclusions}

We have described the production of SPH/SPD from the IR-excitation and photodissociation of molecular beams, with the goal of production rates that approach the IR-photon production rates of ${\rm 10^{21} \, photons \, s^{-1}}$ as close as possible, which depends on the efficiency of the IR-excitation, UV-photodissociation, and target-delivery steps. These proposals are meant to motivate future experimental work, to investigate whether the production rates of conventional SPH production methods can be surpassed significantly, as indicated here. Finally, future expected improvements in the power of IR, visible, and UV lasers will allow further increases in the production rate of this method.

\section*{Acknowledgement}

This work is dedicated to Oleg Vasyutinskii, whose work on polarization in molecular physics has served as an inspiration to us. The authors would also like to gratefully acknowledge Alex Brown for kindly providing the Potential Energy Curves for HCl and HI molecules, as well as his support and advice on the wavepacket propagation dynamics.

\section*{Funding}

This work is supported in part by the Hellenic Foundation for Research and Innovation (HFRI) and the General Secretariat for Research and Technology (GSRT), through the grant agreement No. HFRI-FM17-3709 (project NUPOL).





\end{document}